\documentstyle[aps]{revtex}

\begin{document}
\draft

\author{William J. Mullin}
\address{Department of Physics and Astronomy,\\ University of
Massachusetts, Amherst, MA 01003}

\title{The Loop-Gas Approach to Bose-Einstein Condensation for Trapped Particles}
\date{January 22, 1999 }
\maketitle

\begin{abstract}
We examine Bose-Einstein condensation (BEC) for particles trapped in a
harmonic potential by considering it as a transition in the length of
permutation cycles that arise from wave-function symmetry. This ``loop-gas''
approach was originally developed by Feynman in his path-integral study of
BEC for an homogeneous gas in a box. For the harmonic oscillator potential
it is possible to treat the ideal gas exactly so that one can easily see how
standard approximations become more accurate in the thermodynamic limit
(TDL). One clearly sees that the condensate is made up of very long
permutation loops whose length fluctuates ever more wildly as the number of
particles increases. In the TDL, the WKB approximation, equivalent to the
standard approach to BEC, becomes precise for the non-condensate; however,
this approximation neglects completely the long cycles that make up the
condensate. We examine the exact form for the density matrix for the system
and show how it describes the condensate and behaves in the TDL.
\end{abstract}

\bigskip
\section{INTRODUCTION}

In recent years various groups have achieved Bose-Einstein 
condensation (BEC) in alkali gases.\cite{BEC} The gases have been 
trapped magnetically in potentials accurately approximated by 
harmonic-oscillator wells.  Because of these experiments there has 
been a deluge of theoretical papers on BEC in harmonic potentials, 
both for the ideal and the interacting gases.\cite {String} However, 
even before the recent experimental achievements, there were a few 
theoretical considerations of bosons in harmonic potentials\cite
{OldTheory}$^{-}$\cite{Klep} including one by the author in this journal.%
\cite{WJM1} AJP has recently published two papers on BEC in trapped gases.%
\cite{Ike}$^{,}$\cite{Numer} One of the pleasant aspects of this system is
the relative ease with which substantial theoretical advances can be made
with this system, especially since the diluteness of the experimental gases
often makes the ideal gas a fairly good approximation. The result has been
some new insights into the nature of BEC.

BEC occurs most fundamentally because of the symmetry of the wave 
function.  This property is necessary because bosons are identical 
particles, so that one cannot tell the difference between the 
arrangements, say, $(1,2)$ and $(2,1)$.  This means that forms of 
the wave function that have the particles in differing permutation 
arrangements are equally likely.  For example, for an ideal gas of 
just three particles in single-particle states $\phi _{i}$ might have, 
as one possible state,
\begin{eqnarray}
\psi _{abc}(r_{1},r_{2},r_{3}) &=&\frac{1}{\sqrt{3!}}\left[ \phi
_{a}(r_{1})\phi _{b}(r_{2})\phi _{c}(r_{3})+\phi _{a}(r_{2})\phi
_{b}(r_{1})\phi _{c}(r_{3})\right.  \nonumber \\
&&+\phi _{a}(r_{1})\phi _{b}(r_{3})\phi _{c}(r_{2})+\phi _{a}(r_{3})\phi
_{b}(r_{2})\phi _{c}(r_{1})  \nonumber \\
&&+\left. \phi _{a}(r_{2})\phi _{b}(r_{3})\phi _{c}(r_{1})+\phi
_{a}(r_{3})\phi _{b}(r_{1})\phi _{c}(r_{2})\right]
\end{eqnarray}
Let us call the first term of this function the direct term; the next three
are pair-exchange terms, with just two particles interchanged; the last two
involve triple exchanges with $(1,2,3)\longrightarrow (2,3,1)$ or $(3,1,2)$.
We might represent this by a triangle with lines connecting the 
particles that permute and, because of this picture, call
this a permutation ``cycle'' or ``loop.'' (A pair exchange is a special case
of a loop although it looks like a line.) Any permutation of a set of
particles can be broken down into a combination of loops and singlets
(particles that are not exchanged, e.g., particle 3 in the second term
above).

Feynman developed an approach to Bose systems when he adapted his 
path-integral view of quantum mechanics to statistical mechanics.\cite 
{FeynSM} His work led to new understanding of superfluid helium and of 
the $ \lambda $- transition.\cite{FeynSM}$^{,}$\cite{FeynHe4} The 
path-integral approach involves the use of the configuration 
representation, $\rho(\mathbf{%
r}_{1},...,\mathbf{r}_{N};\mathbf{r}_{1}^{\prime },...,\mathbf{r}%
_{N}^{\prime }),$ of the density matrix as a central entity, a form 
often particularly useful in providing physical insight.  For example, 
the diagonal element gives the probability of finding the $N$ 
particles at positions $ \mathbf{r}_{1},...,\mathbf{r}_{N}.$ In the 
case of bosons, the symmetry of the wave function requires that the 
many-body density matrix include the effects of all possible 
permutations of the $N$ particles.  Since the density matrix is 
bilinear in the wave functions possible for the system, it includes 
not only direct terms (first term squared in our three-particle 
example wave function), but also cross terms among various permutation 
arrangements.  Such cross terms become important when the thermal de 
Broglie wavelength is longer than the separation between particles so 
that particle identity becomes a serious issue; that is, there is a 
substantial overlap between an arrangement and a permutation of that 
arrangement.  

One might expect that small-scale (say, triple) exchange processes (as 
in the product of the first and fifth terms in the above wave 
function) could occur relatively easily when three particles chance to 
be near one another.  What is not so clear is under what circumstances 
very large-scale (say, 50-particle) exchanges can occur as a unit.  
The occurrence of such grand ``ring-around-the-rosy'' exchange loops 
(the overlap of the direct and large-scale exchange terms in the wave 
function) is precisely what happens after the onset of BEC. This view 
is quite different from the usual one of an avalanche of particles 
falling into the lowest state -- although that view is also valid.  At 
BEC there is an ``explosion'' in the size of possible 
permutation-cycle events.  Because their possibility is so clearly a 
result of particle identity -- more evidently than the avalanche into 
the same state -- an analysis of this point of view is a fruitful 
undertaking for the new intuition it provides.  Finding this 
``loop-gas'' relation between BEC and the growth of possible 
permutation cycles is the focus of this paper.

The equilibrium properties of a gas, including the onset of BEC, can 
be determined from the partition function, which is the sum or 
integral over the diagonal terms of the density matrix.  Thus the 
partition function and quantities derived from it can show the effects 
of permutation loops.  Although our result that BEC is accompanied by 
a large increase in the possible size of permutation loops is itself 
not so new; what we show is that considering the partition function 
for particles in a harmonic-oscillator potential allows a particularly 
clear demonstration of the interpretation in terms of permutation 
loops.

In his book, Feynman\cite{FeynSM} gives two derivations of the 
partition function of the homogeneous boson ideal gas in a box with 
periodic boundary conditions.  The first is standard (p.32 of his 
text); the second, arising from his interest in the density matrix 
approach, involves counting all possible permutation loops to compute 
the density matrix and then taking the trace to get the partition 
function (p.62).  (This approach was developed earlier by 
Matsubara,\cite{Matsub} although this article is not referenced by 
Feynman.)  The latter method accurately gives the partition function 
above the transition, but fails, in the homogeneous gas treated, to 
give all the insight for which one might hope, because an 
approximation that must be made to do the needed energy sums neglects 
the effects of the very long permutations.  On the other hand, our 
exact treatment of the ideal gas in a harmonic potential is able to 
avoid this unfortunate approximation, include the condensate 
permutations, and thereby provide a rather useful bit of missing 
pedagogy concerning BEC.

Because we present an explicit computation of the density of the gas 
(from our calculation of the density matrix), we are able to 
demonstrate a curious feature of BEC in a trap.  The density is, of 
course, largest at the center of the trap.  As the temperature is 
lowered, BEC finally occurs when the density at the center reaches the 
\emph{same} critical value that it has for a homogeneous gas with 
periodic boundary conditions.  From the loop-gas point of view, one 
sees that, at this density and temperature, relatively short 
permutation loops suddenly join to make large loops involving 
macroscopic numbers of particles.  In a box, these loops are spread 
out uniformly.  However, in the trap, the density is critical near the 
center, so the macroscopic loops occur in that region.  Of course, 
that feature corresponds precisely to the fact that the ground state 
of the oscillator, into which condensation is taking place, is more 
localized than any other state.

The density matrix itself provides several useful criteria for the
occurrence of BEC, even when interactions are present. Certain integrals
over the density matrix must be macroscopic (of order $N)$, for example.\cite
{PenOns} For a homogeneous fluid of particles in a box with periodic
boundary conditions, the one-body reduced matrix, $\varrho _{1}(\mathbf{r},%
\mathbf{r}^{\prime })$, which is the $N$-body matrix traced over all but the
variables corresponding to a single particle, has the property\cite{PenOns} 
\begin{equation}
\varrho _{1}(\mathbf{r},\mathbf{r}^{\prime })\rightarrow n_{0}/V\,\,\,\,%
\mathrm{for}\,\,\,|\mathbf{r-r}^{\prime }|\rightarrow \infty   \label{rholim}
\end{equation}
where $n_{0}$ is the condensate number and $V$ is the volume. This property
has been useful in Monte Carlo simulations\cite{MCsim} of boson systems,
e.g., superfluid $^{4}$He, for finding $n_{0}$ theoretically. However, if
the particles are in a trap, such as a harmonic well, then $\varrho _{1}(%
\mathbf{r},\mathbf{r}^{\prime })$ will approach zero, simply because of the
nature of the potential, when $|\mathbf{r-r}^{\prime }|\rightarrow \infty $,
and other criteria must then be found. For actual experiments in a trap, the
condensate is recognized by its sudden appearance as a compact cloud of
particles that forms at the center of the trap.

By looking in detail at what is going on in configuration space and at
permutation cycles, rather than by concentrating on the usual view of BEC as
a collapse of many particles into the lowest state, one is able to arrive at
quite different physical insights. Indeed such insights, first developed by
Matsubara and Feynman, have had important applications in current research.

In recent years Ceperley and co-workers have developed a powerful 
numerical procedure, based on Feynman's path integral approach, known 
as Path-Integral Monte Carlo (PIMC).\cite{Ceperley} The technique 
works well for dense systems such as superfluid helium,\cite{Ceperley} 
and for interacting dilute gases as 
well.\cite{Krauth}$^{,}$\cite{Gruter}$^{,}$\cite{HeinMul} The one-body 
density matrix limit above is no help for the trapped gases so that in 
recent PIMC simulations\cite{Krauth}$^{,}$\cite{HeinMul} $n_{0}$ has 
been inferred by looking for the number of particles involved in long 
permutation cycles.  These turn out, not coincidentally, to be 
localized near the center of the trap and, it is argued, constitute 
the condensate.  In the ideal-gas analysis of this paper, such a 
useful connection between long permutation cycles and the localized 
condensate will arise in a simple analytic way without the 
complications of the numerical simulations needed when interactions 
are present.  (Although PIMC is somewhat related to the present paper, 
it is not a focus of this paper and its approach will not be 
investigated.  For more information on it we refer the reader to the 
Ceperley review.\cite {Ceperley})

In another recent related treatment\cite{CentDens} of the dilute Bose, 
all orders of permutation cycles are summed analytically, but while 
also including the effects of interactions in low order.

\section{STANDARD TREATMENT}

We first consider a standard textbook-type treatment of BEC in a harmonic
well -- even though no textbook actually covers this at present. The
harmonic potential we use is written 
\begin{equation}
U(\mathbf{r})=\frac{1}{2}U_{0}\left( \frac{r}{R}\right) ^{2}  \label{Pot}
\end{equation}
where $r$ is the radial distance from the center of the potential, $U_{0}$
is the strength of the potential, and we have included a scale factor $R$
that will become useful later. The energy levels are given in Cartesian
coordinates as 
\begin{equation}
E_{m_{x},m_{y},m_{z}}=\hbar \omega (m_{x}+m_{y}+m_{z}+\frac{3}{2})
\label{ener}
\end{equation}
with $m_{x},m_{y},m_{z}=0,1,2,3,...$ The angular frequency $\omega $ in this
equation is related to the potential parameters by 
\begin{equation}
\omega =\sqrt{\frac{U_{0}}{R^{2}m}}  \label{omeg}
\end{equation}
where $m$ is the particle mass.

We find the properties of the BEC by writing the average particle number in
the grand canonical ensemble 
\begin{equation}
N=\sum_{m_{x},m_{y},m_{z}}\frac{1}{e^{\beta [\hbar \omega (m_{x}+m_{y}+m_{z}+%
\frac{3}{2})-\mu ]}-1}  \label{N}
\end{equation}
where $\beta $ is the inverse temperature, and $\mu $ the chemical potential.

The potential is macroscopic in size -- all $N$ particles fit in it, of
course -- but to take the thermodynamic limit (TDL) we need to increase the
number $N$ while at the same time keeping the average density constant. The
way we do that here is to $\emph{weaken}$ the potential while increasing $N$%
. We now use the scale factor in Eq.(\ref{omeg}); that is, we assume that $%
N/R^{3}\sim N\omega ^{3}$ is to be kept constant while we increase 
$N$.\cite{note1} Further discussion of the TDL that comes to the same 
conclusion is considered in Refs.  4, 7, 21, and 22.  The last 
reference justifies this feature by showing that it is necessary in 
order to keep the free energy extensive.  An equivalent result arises 
in our treatment below.  We can define a characteristic temperature 
$T_{0}$ that remains constant in the TDL by writing
\begin{equation}
\hbar \omega =\hbar \sqrt{\frac{U_{0}}{R^{2}m}}=\hbar \sqrt{\frac{U_{0}}{m}}%
\left( \frac{d}{N}\right) ^{1/3}=\frac{kT_{0}}{N^{1/3}}  \label{omeg2}
\end{equation}
where $d=N/R^{3}$, $k$ is the Boltzmann constant. From Eq.(\ref{omeg2}), we
see that $T_{0}$ is given by 
\begin{equation}
T_{0}=\frac{\hbar }{k}\sqrt{\frac{U_{0}}{m}}d^{1/3}  \label{T0}
\end{equation}

From Eq.(\ref{omeg2}), we see that the harmonic oscillator states are very
closely spaced for large $N$ so that one would expect that it would be a
good approximation to replace the sums in Eq.(\ref{N}) by integrals. We
write 
\begin{eqnarray}
N &=&n_{0}+\int_{0}^{\infty }dm_{x}\int_{0}^{\infty }dm_{x}\int_{0}^{\infty
}dm_{x}\frac{1}{e^{\beta [\hbar \omega (m_{x}+m_{y}+m_{z}+\frac{3}{2})-\mu
]}-1}  \nonumber \\
&=&n_{0}+\left( \frac{kT}{\hbar \omega }\right) ^{3}\int_{0}^{\infty
}du\int_{0}^{\infty }dv\int_{0}^{\infty }dw\frac{1}{e^{u+v+w+\alpha }-1}
\label{varX}
\end{eqnarray}
where we have let $u=\beta \hbar \omega m_{x}$, etc. and 
\begin{equation}
\alpha =\frac{3}{2}\beta \hbar \omega -\beta \mu .  \label{alpha}
\end{equation}
As usual in Bose systems the summand in Eq.(\ref{N}) does not vary uniformly
near $z=u+v+w=0$ (it can peak very sharply there), and so the integral
approximation miscounts $n_{0},$ which becomes macroscopic. Thus we have had
to include it as a special term in Eq.(\ref{varX}). We next expand the
integrand in the usual way and integrate over the three variables. 
\begin{eqnarray}
N &=&n_{0}+\left( \frac{kT}{\hbar \omega }\right) ^{3}\int_{0}^{\infty
}du\int_{0}^{\infty }dv\int_{0}^{\infty }dwe^{-(u+v+w+\alpha )}\frac{1}{%
1-e^{-(u+v+w+\alpha )}}  \nonumber \\
&=&n_{0}+\left( \frac{kT}{\hbar \omega }\right) ^{3}\int_{0}^{\infty
}du\int_{0}^{\infty }dv\int_{0}^{\infty }dw\sum_{l=1}^{\infty
}e^{-l(u+v+w+\alpha )}  \nonumber \\
&=&n_{0}+\left( \frac{kT}{\hbar \omega }\right) ^{3}\sum_{l=1}^{\infty
}e^{-l\alpha }\left( \int_{0}^{\infty }due^{-lu}\right) ^{3}=n_{0}+\left( 
\frac{kT}{\hbar \omega }\right) ^{3}F_{3}(\alpha )  \label{uvwint}
\end{eqnarray}
where 
\begin{equation}
F_{3}(\alpha )=\sum_{l=1}^{\infty }e^{-l\alpha }\frac{1}{l^{3}}  \label{F3}
\end{equation}
is one of the Bose integrals.\cite{Boseint}

If we use Eq.(\ref{omeg2}), we find 
\begin{equation}
N=n_{0}+N\left( \frac{T}{T_{0}}\right) ^{3}F_{3}(\alpha ).  \label{N2}
\end{equation}
From this equation we see that our procedure for taking the TDL has led us
to a properly extensive expression for the non-condensed particles. Any
relation between $\omega $ and $N$ other than $\omega ^{3}\sim 1/N$ would
not lead to this result. For non-zero $\alpha $ , we can satisfy this
equation with a negligible value of $n_{0}.$ However, as $T$ falls, the value
of $\alpha $ needed to satisfy the equation approaches zero (becomes of
order $1/N$) and $F_{3}$ reaches its maximum value of $\zeta (3)$, where $%
\zeta (n)$ is a Riemann zeta-function. Bose-Einstein condensation takes
place at this point and we have 
\begin{equation}
n_{0}=N\left[ 1-\left( \frac{T}{T_{c}}\right) ^{3}\right]
\,\,\,\,\,\,\,\,\,\,\,\mathrm{for}\,\,\,\,\,\,\,T<T_{c}  \label{n0}
\end{equation}
where $T_{c}=T_{0}\zeta (3)^{-1/3}$ is the condensation temperature. In Fig.
1 we plot the condensate fraction $n_{0}/N$ as a function of $T$ as gotten
from Eq.(\ref{n0}), and also from the exact expression Eq.(\ref{N})
evaluated for finite $N=1000$.\cite{detail}

Eq.(\ref{F3}) contains a sum over $l$. Just what is the physical
significance of that sum? For the above derivation it resulted only from a
mathematical trick invoked to do an integral, but in fact it has a
fundamental meaning in terms of permutation loops as we will show in Sec. IV.

There is a considerably different looking derivation that has been used in
the literature for this problem.\cite{Schick}$^{,}$\cite{Klep} It involves
the WKB, or semiclassical, approximation for the density $n(\mathbf{p},%
\mathbf{r})$ of particles having momentum $\mathbf{p}$ and position $\mathbf{%
r}$, which is 
\begin{equation}
n(\mathbf{p},\mathbf{r})=\frac{1}{h^{3}}\frac{1}{e^{\beta [\mathbf{p}%
^{2}/2m+U(\mathbf{r})-\mu ]}-1}  \label{npr}
\end{equation}
The density of particles at position $\mathbf{r}$ is the integral of this
over $\mathbf{p,}$ which can be shown, by an expansion similar to that done
above, to be 
\begin{eqnarray}
\rho (\mathbf{r}) &=&\frac{1}{h^{3}}\int d\mathbf{p}\frac{1}{e^{\beta [%
\mathbf{p}^{2}/2m+U(\mathbf{r})-\mu ]}-1}  \nonumber \\
&=&\left( \frac{2\pi mkT}{h^{2}}\right) ^{3/2}\sum_{l=1}^{\infty }\frac{%
e^{l\beta \mu }}{l^{3/2}}e^{-\beta lU(\mathbf{r})}  \label{WKBrhor}
\end{eqnarray}
If we integrate this over $\mathbf{r}$ with the quadratic potential of Eq.(%
\ref{Pot}), we will find just the second term on the right-hand side of Eq.(%
\ref{N2}).  That is, the WKB approximation does not include the 
condensate, which must be added on specially as in our previous 
derivation.

We have given the standard approach to BEC for the trapped gas case. In the
remaining sections we examine how we can get considerably more insight into
BEC by looking at the process from the point of view of permutation loops;
that is, we examine the so-called ``loop gas.'' But first we consider an
exact approach that avoids replacing sums by integrals.

\section{AN EXACT APPROACH}

When we replaced the sums by integrals above to get an answer that would be
applicable in the thermodynamic limit, we lost the correct description of
the condensate, which was added back in. In the case of the harmonically
trapped gas, we can avoid this approximation by doing the sums exactly. In
Eq.(\ref{N}) we make the expansion of the integrand as done in Eq.(\ref
{uvwint}): 
\begin{equation}
N=\sum_{m_{x},m_{y},m_{z}}\sum_{l=1}^{\infty }e^{-l\beta \hbar \omega
(m_{x}+m_{y}+m_{z})-l\alpha }  \label{nn}
\end{equation}
Unlike the case of a free gas in a box where the energies go quadratically
with the quantum numbers, the sums over $m_{x},m_{y},$ and $m_{z}$ can be
done \emph{exactly}. The result is 
\begin{equation}
N=\sum_{l=1}^{\infty }\frac{e^{-\alpha l}}{(1-e^{-l\beta \hbar \omega })^{3}}
\label{NNN}
\end{equation}
Such expressions have been obtained in the literature previously.\cite
{Others}

To retrieve the approximations of Sec. I, we need to note that $\omega $ is
of order $1/N^{1/3}$ so that, for $l$ smaller than $N^{1/3}$, a very large
number in the TDL, the argument of the exponential in the denominator is
small. Expand in powers of it and keep only the first term of the expansion
to find 
\begin{equation}
N^{\prime }=\sum_{l=1}^{\infty }\frac{e^{-\alpha l}}{(l\beta \hbar \omega
)^{3}}=\left( \frac{kT}{\hbar \omega }\right) ^{3}F_{3}(\alpha )
\label{Approx}
\end{equation}
This is identically the same expression found in the last section for the
non-condensed gas. The condensate has been neglected because $\alpha $ is of
order $1/N$ for $T<T_{c}$ as can be seen from the expression for $n_{0}$ : 
\begin{equation}
n_{0}=\frac{1}{e^{\alpha }-1}=\sum_{l=1}^{\infty }e^{-\alpha l}\cong \frac{1%
}{\alpha }  \label{n00}
\end{equation}
where the last approximation holds only below $T_{c}$.  The size of 
$\alpha $ means that the $l$ sum is significant to values up to order 
$N$.  Since the approximation we used to get Eq.(\ref{Approx}) was 
valid only for $l<N^{1/3}$, it is no surprise that we have made an error in 
deriving Eq.(\ref{Approx}).  We can see how this occurs in Fig.  2, 
which plots as a function of $l$, for finite $N=1000,$ the exact 
summand of Eq.(\ref{NNN}), the approximate summand of 
Eq.(\ref{Approx}), and the summand of $n_{0}$ of Eq.(\ref{n00}).  (The 
last two summands do not quite add up to the exact total because of 
higher order terms neglected in Eq.(\ref{Approx}).  These would become 
smaller in the TDL.) The fact that the standard integration 
approximation neglects large $l$ terms again draws attention to the 
question of the physical meaning of these terms.  In the next section 
we show how the sum represents permutations cycles.

\section{LOOP GAS APPROACH}

The ``loop gas'' view was first introduced by Matsubara\cite{Matsub} and
then later by Feynman in connection with his path-integral method in
statistical mechanics\cite{FeynSM}; it has also been discussed by Elser\cite
{Elser}, who seems to have invented the terminololgy ``loop gas.'' These
references all deal with the homogeneous system with periodic box boundaries.
The term ``loops'' is used here as shorthand for ``permutation cycles'' or
``exchange cycles.'' The reader will find that the derivation given here
is more complicated than that of the last two sections, but the motivation
is to achieve new physical insight into the relation between BEC and the
growth of permutation cycles.

The boson $N$-body density matrix involves only symmetric states and is\cite
{FeynSM} 
\begin{equation}
\rho _{S}(\mathbf{r}_{1},...\mathbf{r}_{N};\mathbf{r}_{1}^{\prime },...,%
\mathbf{r}_{N}^{\prime })=\frac{1}{N!}\sum_{P}\rho _{D}(\mathbf{r}%
_{P_{1}},...\mathbf{r}_{P_{N}};\mathbf{r}_{1}^{\prime },...,\mathbf{r}%
_{N}^{\prime })  \label{Ndens}
\end{equation}
where the sum is over all permutations and 
\begin{equation}
\rho _{D}(\mathbf{r}_{1},...\mathbf{r}_{N};\mathbf{r}_{1}^{\prime },...,%
\mathbf{r}_{N}^{\prime })=\sum_{\mathrm{all\,\,\,states}}e^{-\beta
E_{i}}\psi _{i}(\mathbf{r}_{1},...,\mathbf{r}_{N})\psi _{i}^{*}(\mathbf{r}%
_{1}^{\prime },...,\mathbf{r}_{N}^{\prime })~
\end{equation}
is the density matrix for distinguishable particles. The partition function
is the trace of this matrix or 
\begin{equation}
Z=\frac{1}{N!}\sum_{P}\int d\mathbf{r}_{1}...d\mathbf{r}_{N}\langle \mathbf{r%
}_{P_{1}},...,\mathbf{r}_{P_{N}}|e^{-\beta H}|\mathbf{r}_{1},...,\mathbf{r}%
_{N}\rangle
\end{equation}
where the variable $\mathbf{r}_{Pj}$ represents the coordinate of the
particle interchanged with particle $j$ in permutation $P.$ The sum over
permutations here is just the result of the possible cross terms between
various permuted elements of the symmetrized wave functions caused by
particle identity. By using the configuration representation of the density
matrix, we are able to track how the symmetrization affects the results.

Any $N$-particle permutation breaks up into a number of smaller loop
permutations. For example, for $N=7$ we might have a 4-particle loop, a pair
permutation (2-particle loop), and a single particle. In the 4-particle
loop, we might have particle \#1 interchanged with \#2, \#2 interchanged
with \#3, \#3 interchanged with \#4, and \#4 with \#1. Given a set of
configurations containing all $C(q_{1},q_{2},...)$ ways of having $q_{1}$
loops of length 1, $q_{2}$ loops of length 2, etc$.,$ we can write 
\begin{equation}
Z_{N}=\frac{1}{N!}\sum_{\{q_{1},q_{2}...\}}C(q_{1},q_{2},...)%
\prod_{l}Z(l)^{q_{l}}  \label{ZN}
\end{equation}
where the sum is over all combinations of permutation numbers such that 
\begin{equation}
\sum_{l}q_{l}l=N,  \label{restrict}
\end{equation}
and 
\begin{equation}
Z(l)=\int d\mathbf{r}_{1}...d\mathbf{r}_{l}\langle \mathbf{r}_{P_{1}},...,%
\mathbf{r}_{P_{l}}|e^{-\beta H}|\mathbf{r}_{1},...,\mathbf{r}_{l}\rangle 
\end{equation}
corresponds to a unbroken permutation loop of $l$ particles. Feynman\cite
{FeynSM} has given an argument to show that this quantity is 
\begin{equation}
C(q_{1},q_{2},...)=\frac{N!}{1^{q_{1}}2^{q_{2}}...q_{1}!q_{2}!...}
\label{count}
\end{equation}
While the argument given by Feynman can be consulted, for completeness we
rederive this result in the Appendix. Eq.(\ref{ZN}), with the result Eq.(\ref
{count}), is quoted in a beautiful little paper on Bose and Fermi partition
functions by D. I. Ford\cite{Ford}, as well as by Matsubara and Feynman.

To evaluate the most probable distribution of loops, we invoke the grand
canonical ensemble and multiply $Z_{N}$ by $e^{\beta \mu N}$ and sum over
all particle $N$ values. This trick removes the restriction of Eq.(\ref
{restrict}) and allows the individual sums on $q_{l}$ to be done. The grand
canonical partition function is 
\begin{eqnarray}
\mathcal{Z} &=&\sum_{N}e^{\beta \mu N}Z_{N}=\prod_{l}\sum_{q_{l}}%
\frac{1}{q_{l}!}\left( \frac{Z(l)e^{\beta \mu l}}{l}\right) ^{q_{l}} 
\nonumber \\
&=&\prod_{l}\exp \left\{ \frac{Z(l)e^{\beta \mu l}}{l}\right\} 
\end{eqnarray}
What we want from this is the average particle number, $\left\langle
N\right\rangle ,$ and the average number of particles, $\left\langle
q_{l}\right\rangle ,$ in the permutation loop of length $l.$ We use the
standard relation 
\begin{equation}
\left\langle N\right\rangle =\frac{\partial }{\partial \mu }kT\ln \mathcal{Z}%
=\sum_{l}Z(l)e^{\beta \mu l}  \label{Nave}
\end{equation}
From this it is clear that the average value of the loop number $q_{l}$ is 
\begin{equation}
\left\langle q_{l}\right\rangle =\frac{Z(l)e^{\beta \mu l}}{l}  \label{nlave}
\end{equation}

Because $H$ is the Hamiltonian corresponding to an ideal gas, $Z(l)$ can be
broken up into a product of single-particle densities. $Z(l)$ is most easily
evaluated by making each single-particle operator $e^{-\beta H_{i}}$
diagonal by interposing complete sets of harmonic oscillator functions as
follows: 
\begin{eqnarray}
Z(l) &=&\sum_{m_{1},m_{2},...}\int d\mathbf{r}_{1}...d\mathbf{r}%
_{l}\left\langle \mathbf{r}_{P_{1}},...,\mathbf{r}_{P_{l}}|m_{1},...,m_{l}%
\right\rangle ~~~~~~  \nonumber \\
&&\times e^{-\beta (E_{m_{1}}+...+E_{m_{l}})}\left\langle m_{1},...,m_{l}|%
\mathbf{r}_{1},...,\mathbf{r}_{l}\right\rangle 
\end{eqnarray}
where we have included only one subscript $m_{i}$ to represent all three
quantum numbers for a single-particle state. The $\mathbf{r}$ integrations
lead to factors like 
\begin{equation}
\int d\mathbf{r}_{1}\left\langle m_{P_{k}}|\mathbf{r}_{1}\right\rangle
\left\langle \mathbf{r}_{1}|m_{1}\right\rangle =\delta _{m_{P_{k}},m_{1}}
\end{equation}
where $P_{k}$ represents the particle number that moved into position $%
\mathbf{r}_{1}$ in the permutation. Since we have a ring exchange these $%
\delta $-functions mean that \emph{all} $m_{i}$ become equal to one of them,
call it $m.$ The result is simply 
\begin{equation}
Z(l)=\sum_{m}e^{-\beta lE_{m}}  \label{ZL}
\end{equation}
which is a one-body partition function at inverse temperature $\beta l.$ The
sum, over the three quantum numbers represented by $m$ is easily done: 
\begin{equation}
Z(l)=\frac{e^{-3\beta l\hbar \omega /2}}{\left( 1-e^{-\beta l\hbar \omega
}\right) ^{3}}
\end{equation}
The numerator comes from the zero point energy in each $E_{m}$. Putting this
result back into Eq.(\ref{Nave}) (and dropping the averaging brackets), we
find 
\begin{equation}
N=\sum_{l}\frac{e^{-\alpha l}}{\left( 1-e^{-\beta l\hbar \omega }\right) ^{3}%
}  \label{Nagain}
\end{equation}
which, as expected, is precisely the same result as derived in Sec.  
III. However, this derivation has led to the very instructive result 
that the sum over $l$ is a sum over \emph{permutation loops}, which 
arise from the symmetrization of the wave function.  Note also that in 
the homogenous gas in a box, one is unable to do the exact sum over 
states as we could in Eq.(\ref {ZL}).  In that 
case\cite{FeynSM}$^{,}$\cite{Matsub}$^{,}$\cite{Elser} one replaces 
the sum by an integral; the result is that the expression equivalent 
to Eq.(\ref{Nagain}%
) is approximate and neglects the condensate, which must once more be added
in by hand. Eq.(\ref{Nagain}), on the other hand, is exact and \emph{includes%
} the condensate.

Figure 2 illustrates the fact that the condensate is made up of 
large-scale permutation loops, a fact that is well-known, but has not 
been so well-illustrated previously.  What has perhaps not been so 
clear previously is that loops of \emph{all} sizes contribute nearly 
equally to the condensate. The average number of particles 
$p_{l}=\left\langle q_{l}\right\rangle l$ in a permutation loop is 
given by Eq.(\ref{nlave}) as
\begin{equation}
p_{l}=\frac{e^{-\alpha l}}{(1-e^{-lT_{0}/TN^{1/3}})^{3}}
\end{equation}
where we have used Eq.(\ref{omeg2}). Because $\alpha $ is of order $1/N$ at $%
T<T_{c},$ the $l$ sum extends to order $N,$ but we know that the
non-condensed particles are given by terms with $lT_{0}/TN^{1/3}\ll 1,$ that
is, for loops of size less than of order $N^{1/3}.$ The terms for larger $%
l\gg N^{1/3}$ have $p_{l}\cong e^{-\alpha l},$ which is identical with 
the form given in the second last version of Eq.(\ref{n00}) for 
$n_{0}.$ All these terms then contribute only to the condensate.  Each 
such term in the condensate is crudely of order 1, but there are of 
order $N$ such terms so they contribute a total number of particles on 
the order of $N.$ The fact that permutation loops of 
length of order $l>N^{1/3}$ have only 1 particle in them on 
average means that the long permutation loops are 
forming and breaking with fluctuations in size that are extremely 
wild.\cite{note2}

\section{DENSITY MATRIX}

The density matrix is a very useful quantity allowing averages of any
thermodynamic variable to be evaluated. The one-body reduced density,
mentioned in Sec. I, is particularly useful in considerations of BEC as
Penrose and Onsager\cite{PenOns} have shown. This quantity is defined by
integrating the $N$-body density, Eq. (\ref{Ndens}) over all but one
coordinate: 
\begin{equation}
\varrho _{1}(\mathbf{r},\mathbf{r}^{\prime })=\int d\mathbf{r}_{2}...d%
\mathbf{r}_{N}~\rho _{S}(\mathbf{r},\mathbf{r}_{2}...\mathbf{r}_{N};\mathbf{r%
}^{\prime },\mathbf{r}_{2}...\mathbf{r}_{N})
\end{equation}
The diagonal element $\varrho _{1}(\mathbf{r},\mathbf{r})$ is just the
density of particles $\varrho (\mathbf{r})$ at position $\mathbf{r}$. It
turns out, after a short derivation, that the one-body reduced density for
the boson ideal gas has the rather simple, physically intuitive form 
\begin{equation}
\varrho _{1}(\mathbf{r},\mathbf{r}^{\prime })=\sum_{m}n_{m}\psi _{m}(\mathbf{%
r})\psi _{m}^{*}(\mathbf{r}^{\prime })  \label{OneBodDen}
\end{equation}
where 
\begin{equation}
n_{m}=\frac{1}{e^{\beta (E_{m}-\mu )}-1}
\end{equation}
is simply the occupation of the $m$th single-particle state as used in Eq.(%
\ref{N}), and $\psi _{m}$ is the corresponding single-particle state.

The density matrix $\varrho _{1}(\mathbf{r},\mathbf{r}^{\prime })$ can be
considered to be an operator with eigenvalues. In fact the spectral form of
Eq.(\ref{OneBodDen}) shows that the eigenfunctions are the $\psi _{m}$ and
the eigenvalues are the $n_{m};$ it is straightforward to verify that $\int d%
\mathbf{r}^{\prime }\varrho _{1}(\mathbf{r},\mathbf{r}^{\prime })\psi _{m}(%
\mathbf{r}^{\prime })=n_{m}\psi _{m}(\mathbf{r}).$ The fact that $\varrho
_{1}(\mathbf{r},\mathbf{r}^{\prime })$ has a \emph{large} eigenvalue, in
this case $n_{0},$ leads to its usefulness as a criterion for BEC. Even when
there are interactions, the large eigenvalue is the condensate number and
the eigenfunction is the condensate ``wave function,'' but note that the
solution is no longer simply the ground-state of the oscillator potential.
In the case of weakly interacting bosons, this condensate wave function is
approximately the solution of the well-known Gross-Pitaevskii nonlinear
Schr\"{o}dinger equation.\cite{String}

We can transform the density into a sum over permutation loops by expanding $%
n_{m}$ in the now familiar way. We get 
\begin{equation}
\varrho _{1}(\mathbf{r},\mathbf{r}^{\prime })=\sum_{l=1}^{\infty }e^{\beta
\mu l}\left\{ \sum_{m}e^{-l\beta E_{m}}\psi _{m}(\mathbf{r})\psi _{m}^{*}(%
\mathbf{r}^{\prime })\right\} 
\end{equation}
The quantity in curly brackets is simply the harmonic oscillator density
matrix $d(\mathbf{r},\mathbf{r}^{\prime })$ for a \emph{single} free
(distinguishable) particle, but at inverse temperature $l\beta ,$ rather
than simply $\beta .$ The value of this is well known, and is\cite{FeynSM} 
\begin{eqnarray}
d_{l}(\mathbf{r},\mathbf{r}^{\prime }) &=&\left( \frac{m\omega }{h\sinh
\left( l\beta \hbar \omega \right) }\right) ^{3/2}  \\
&\times& \prod_{i=x,y.z}\exp \left\{ -\frac{m\omega }{2\hbar \sinh (l\beta
\hbar \omega )}\left[ (r_{i}^{2}+r_{i}^{\prime 2})\cosh (l\beta \hbar \omega
)-2r_{i}r_{i}^{\prime 2}\right] \right\}  \nonumber
\end{eqnarray}
One can see, for example, that $d_{l}(\mathbf{r},0),$ and therefore $\varrho
_{1}(\mathbf{r},0),$ goes to zero at large $\mathbf{r}$ in contrast to the
homogeneous case of Eq.(\ref{rholim}).

We are particularly interested in the particle density $\rho (\mathbf{r}%
)=\varrho _{1}(\mathbf{r},\mathbf{r}).$ With a bit of hyperbolic function
manipulation, one finds this to be 
\begin{equation}
\rho (\mathbf{r})=\sum_{l=1}^{\infty }e^{\beta \mu l}\left( \frac{m\omega }{%
h\sinh \left( l\beta \hbar \omega \right) }\right) ^{3/2}\exp \left\{ -\frac{%
m\omega }{\hbar }\tanh \left( \frac{l\beta \hbar \omega }{2}\right) \mathbf{r%
}^{2}\right\}  \label{denden}
\end{equation}
One can integrate this $\rho (\mathbf{r)}$ to verify the result of Eq.(\ref
{NNN}).

We have seen that the WKB approximation can be found by assuming that $%
l\beta \hbar \omega $ is small. If we do that here we find 
\begin{equation}
\rho (\mathbf{r})=\sum_{l=1}^{\infty }e^{\beta \mu l}\left( \frac{2\pi m}{%
h^{2}l\beta }\right) ^{3/2}\exp \left\{ -\frac{l\beta m\omega ^{2}}{2}%
\mathbf{r}^{2}\right\}
\end{equation}
This result agrees with Eq.(\ref{WKBrhor}) for the harmonic potential.

The largest value of the density occurs at the origin, of course, and we can
ask what this value is at the BEC transition. For $T$ above or equal to the
transition temperature, the WKB approximation is valid in the TDL. Right at
the transition the chemical potential vanishes (actually it is the quantity $%
\alpha $ of Eq.(\ref{alpha}) that vanishes, but the WKB approximation
neglects the zero-point energy, which is of order $1/N^{1/3}$.) We find that
the critical density is 
\begin{equation}
\rho _{c}(0)=\left( \frac{2\pi mkT_{c}}{h^{2}}\right)
^{3/2}\sum_{l=1}^{\infty }\frac{1}{l^{3/2}}=\left( \frac{2\pi mkT_{c}}{h^{2}}%
\right) ^{3/2}\zeta (3/2)
\end{equation}
or 
\begin{equation}
kT_{c}=\frac{2\pi \hbar ^{2}}{m}\left( \frac{\rho _{c}(0)}{\zeta (3/2)}%
\right) ^{2/3}
\end{equation}
But this is precisely the transition temperature for an ideal gas in a \emph{%
box} at \emph{uniform} density $\rho _{c}(0)$! Thus the trapped gas, which
has its BEC localized near the origin, has a condensate there under just the
same conditions as the gas in a box. This interesting situation has been
noted previously in the literature.\cite{Klep}$^{,}$\cite{CentDens}$^{,}$%
\cite{HKN}.

In Fig. 3 we plot the full density as a function of position as well as the
WKB approximation to it. One sees that the two are equal only for large $r.$
In the case of $N=1000$ at the temperature considered, the condensate number
is about 456, so the non-condensed particle number is 544, despite the
appearance in the plot that the condensate is much larger than the
non-condensate. This feature is explained by the fact that there is a factor 
$r^{2}$ involved when one integrates this to find the total $N$ value; this
factor diminishes the interior points and emphasizes the wide wings.

Another way we can see that the high $l$ permutation loops belong to
the condensate is by considering the mean square radius of the individual
permutation densities. This is 
\begin{equation}
\sigma _{l}^{2}=\frac{\int d\mathbf{r}r^{2}\rho _{l}(\mathbf{r})}{\int d%
\mathbf{r}\rho _{l}(\mathbf{r})}
\end{equation}
where $\rho _{l}$ is the summand of Eq.(\ref{denden}). The result is 
\begin{equation}
\sigma _{l}^{2}=\frac{3}{2}\frac{\hbar }{m\omega }\frac{1}{\tanh l\beta
\hbar \omega /2}
\end{equation}
The WKB approximation for this is gotten by considering $l\beta \hbar \omega 
$ small, for which we find 
\begin{equation}
\sigma _{l}^{2}(\mathrm{WKB})=3\frac{\hbar }{m\omega }\frac{1}{l\beta \hbar
\omega }
\end{equation}
This quantity drops off like $1/l$. The contribution from the condensate
comes from large $l\beta \hbar \omega $ ($N^{1/3}<l\leq N$) and is 
\begin{equation}
\sigma _{l}^{2}(\mathrm{condensate})=\frac{3}{2}\frac{\hbar }{m\omega }
\end{equation}
This is simply the mean square width of the ground-state of the harmonic
oscillator. Figure 4 shows the plot of $\sigma _{l}^{2}$, $\sigma _{l}^{2}(%
\mathrm{WKB})$, and $\sigma _{l}^{2}(\mathrm{condensate})$ for $N=1000$. We
again see that, for $l>40$ or so, that the permutation loops all refer to
the condensate.

In recent path-integral Monte Carlo 
calculations\cite{Krauth}$^{,}$\cite {HeinMul} for trapped gases, it 
was necessary to identify the condensate by use of the permutation 
cycles.  Ref.16 assumed that the size of the largest cycle occurring 
was equivalent to the condensate number.\cite{note2} Ref.18, on the 
other hand, assumed that the condensate was made up of all particles 
for $l$ greater than a value where $\sigma _{l}^{2}$ has become 
constant.  However, we see that, for the latter approach, some of the 
particles having \emph{small} permutation cycles also contribute to 
the condensate and must be counted.

\section{DISCUSSION}

We have seen that it is possible and instructive to examine BEC via the
permutation loop picture. Ideal gases trapped in a harmonic potential become
particularly helpful in this regard because we can do many of the
computations analytically and exactly. The very long permutation loops are
characteristic of BEC and form when the thermal de Broglie wavelength becomes
comparable to the average distance between particles. Under these conditions
multiparticle exchange becomes possible and the coherence characteristic of
BEC takes place with the particles exhibiting the strong symptoms of
indistinguishability.

The best known characteristic of BEC is the presence of a macroscopic 
fraction of the particles in the lowest state.  For the ideal gas, in a 
harmonic trap at absolute zero, those particles would be in the 
oscillator ground-state.  From the point of view of permutations, this 
situation is equivalent to all possible lengths of permutation cycles 
being equally likely (up to length $N$, which is taken to infinity in 
the TDL).  As the temperature is raised, but still below the 
condensation temperature, the shorter loops become increasingly 
prevalent.  Shorter loops occur when particle indistinguishability is 
less a factor and corresponds to particles being spread out over 
excited oscillator states.  Finally, above the transition temperature, 
no state has a macroscopic occupation of particles and most 
permutation cycles are short; very long ones are rare.

BEC and the $\lambda $-transition in liquid helium have been compared with a
polymer transition,\cite{Ceperley} with long chain polymers forming below
the transition temperature. In fact there is a transition in sulphur\cite
{sulphur} in which the opposite effect occurs: above the transition there
are very long chains of sulphur atoms, while below the transition only
eight-atom rings form; the isomorphism involves $\beta \rightarrow kT.$
Interestingly the specific heat of sulphur has a ``$\lambda $-transition.''

The reduced single-particle density matrix has the lowest state as an
eigenfunction with the condensate number as eigenvalue. For the ideal gas
demonstrating this is a straightforward matter with its connection to long
permutation loops made evident in the density. However, the property carries
over into the interacting gas as well. Finding the eigenfunction of the
one-particle density is one of the most important theoretical problems in
the interacting fluid, with the solution representing the condensate wave
function and the ``order parameter'' of the state. The solution of the
Gross-Pitaevskii equation is a standard approximation to this wave function.
In the case of real harmonically trapped gases, which actually do interact,
this equation seems to represent the experimental situation fairly
accurately.\cite{String}

The PIMC approach was invented by Feynman and has been exploited in a wide
range of fields, including the description of polymers.\cite{PolymerBk} One
of its most successful applications has been to bosons and liquid $^{4}$He
by Ceperley and co-workers.\cite{Ceperley} The approach considers in detail
the particle positions in space, using the expression Eq.(\ref{Ndens}) as
its basis, and takes the formation of permutation loop ``polymers'' as one
of its most important features. The interested reader should consult the
review of Ceperley.\cite{Ceperley}

\vspace{0.3in} \noindent\textbf{APPENDIX} \vspace{0.2in}

In this Appendix we derive the expression Eq.(\ref{count}) for the number of
ways of forming $q_{1}$ permutation loops of length 1, $q_{2}$ of length 2,
...$q_{l}$ of length $l,$ etc. Suppose we consider an 9-particle system
having two 3-particle loops [(1$\rightarrow 2\rightarrow 3\rightarrow 1)$
and (4$\rightarrow 5\rightarrow 6\rightarrow 4)]$, a 2-particle loop [(7$%
\rightarrow 8\rightarrow 7)]$, and a singlet [(9)]. We can mix up these
particles  to get other non-equivalent 3-, 2-, and 1-loops. We want to count
ways we can do this. For example, suppose we replace 1 by 4 and 4 by 1,
giving (4$\rightarrow 2\rightarrow 3\rightarrow 1),$ (1$\rightarrow
5\rightarrow 6\rightarrow 1),$ $(7\rightarrow 8\rightarrow 7),$ and $(9).$
This is a new arrangement. There are 5! different ways of mixing the
particles to get new permutations arrangements, but not all of them are
distinct. For example if we replace 2 by 1, 3 by 2, and 1 by 3, we get the
permutation loop (3$\rightarrow 1\rightarrow 2\rightarrow $ 3), which is
just the same as the original arrangement, with the loop rotated one click,
which makes no real change. In a loop of length $l$ there are $l$ such loop
rotations that don't count for new arrangements. Another way of replacing
particles that does \emph{not} lead to new arrangements is by interchanging
all the particles in one 3-loop of the same length by those of the other
3-loop. There are 2! such interchanges of loops possible in our special
case. Thus the total number of possible distinct arrangements in our
9-particle case is 9!/3$^{2}2!$. For the N-particle case we easily generalize
to find that the number of distinct arrangements for the $N$ particle case
is 
\begin{equation}
C(q_{1},q_{2},...)=\frac{N!}{1^{q_{1}}2^{q_{2}}...q_{1}!q_{2}!...}
\label{Cauchy}
\end{equation}
where $N!$ is the total number of ways of rearranging $N$ particles, $%
l^{q_{l}}$ is the number of ways of ``rotating'' the $q_{l}$ permutations
loops of length $l,$ none of which produces a new arrangement, and $%
q_{l}!$ is the number of ways of interchanging whole loops of the same
length, which also does not produce a new arrangement. The expression of Eq.(%
\ref{Cauchy}) is known as ``Cauchy's formula.''\cite{Berge}

\vspace{0.3in} \noindent\textbf{ACKNOWLEDGMENTS} \vspace{0.2in}

I am grateful for many useful conversations about BEC with J. P. Fernandez,
Franck Lalo\"{e}, Stefan Heinrichs, Werner Krauth, and Markus Holzmann.

\vspace{0.3in} \noindent\textbf{FIGURE CAPTIONS} \vspace{0.2in}

\noindent \emph{Figure 1}. Condensate fraction versus $T/T_{0}.$ The
condensation transition temperature is $T_{0}/\zeta (3)^{1/3}=$ 0.94$T_{0}.$
The solid line is the thermodynamic limit case given by Eq.(\ref{n0}). The
dotted line is the exact solution for $N=$1000.\medskip

\noindent \emph{Figure 2}. Summands of various expressions of Sec. III for $%
N $ versus summation index $l$. Here we have $N=1000$ and $T/T_{0}=$ 0.7.
The solid line is the summand of the exact expression Eq.(\ref{NNN}); the
dashed line represents the WKB summand of Eq.(\ref{Approx}); and the dotted
line is the summand of the condensate in Eq.(\ref{n00}). The maximum value
of the exact case at $l=1$ is 423. Since the index $l$ corresponds to
permutation loops, that means there are 423 singlets; the condensate (the
dotted line) includes all the large $l$ values (the long permutation loops)
as well as some of the small loops and corresponds to a total of 456
particles; the remaining small, non-condensate loops contribute 121 
particles. Note that the summands of Eq.(\ref{Approx}) and Eq.(\ref{n00}) do
not quite add up to the exact result, because of the higher order neglected
terms in the WKB expression. \medskip

\noindent \emph{Figure 3}. Density as a function of $r.$ The solid line is
the exact density, the dashed line is the WKB approximation, which
represents the non-condensed particles, and the dotted line is the
condensate density, derived using the first term in sum of Eq.(\ref
{OneBodDen}). \medskip $T/T_{0}=0.8$ here. Density is in units of $(m\omega
/\hbar )^{3/2}$ and position in units of $(\hbar /m\omega )^{1/2}.$\medskip

\noindent \emph{Figure 4}. Mean square width (in units of $(\hbar /m\omega
)) $ for particles in permutation loop $l$ versus $l.$ The solid line is the
exact result, the dashed line is the WKB approximation to the mean-square
width, representing the non-condensed particles, and the dotted line is the
condensate width (3/2).


\begin{thebibliography}{99}
\bibitem{note1} Of course, the variable $R$ $\,$is redundant; it would 
be enough simply to maintain $N\omega ^{3}$constant as $N $ is 
increased.  With the convenient introduction of $R$ we assume that 
$U_{0} $ is kept fixed in the TDL.
\bibitem{detail}  For numerical computations, Eq. (\ref{N}) was transformed
into a one-dimensional sum over $p=m_{x}+m_{y}+m_{z}.\ $We found that value
of $\alpha $ that made the sum numerically equal to the appropriate value of 
$N,$ in this case 1000. The value $\alpha $ leads to $n_{0}$ via Eq. (\ref
{n00}).
\bibitem{note2} The size of possible permutation loops extends to $N$ 
in the grand canonical evaluation used here. However, in a true 
canonical ensemble evaluation of $p_{l}$ the largest loop is the size of the 
condensate.\cite{Krauth}
\end{thebibliography}
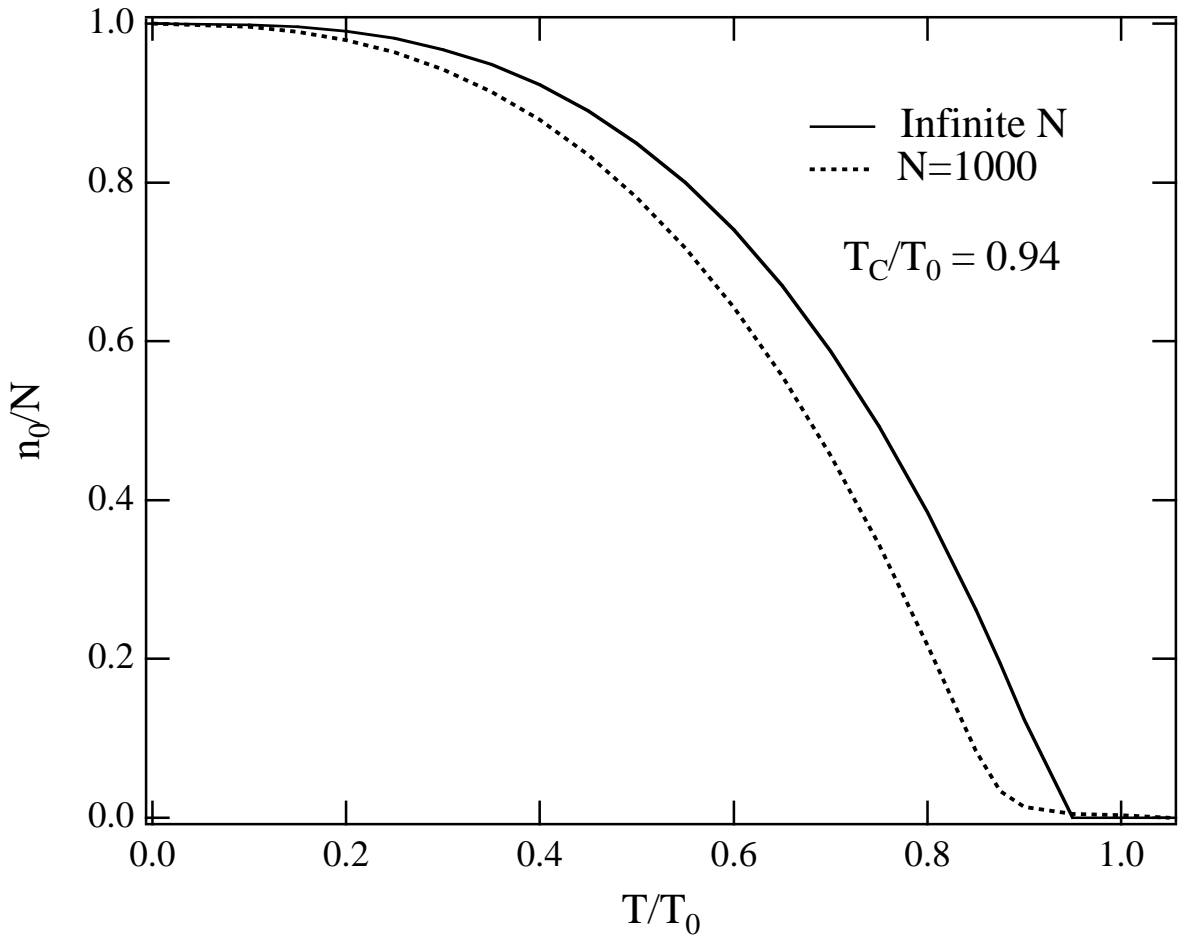
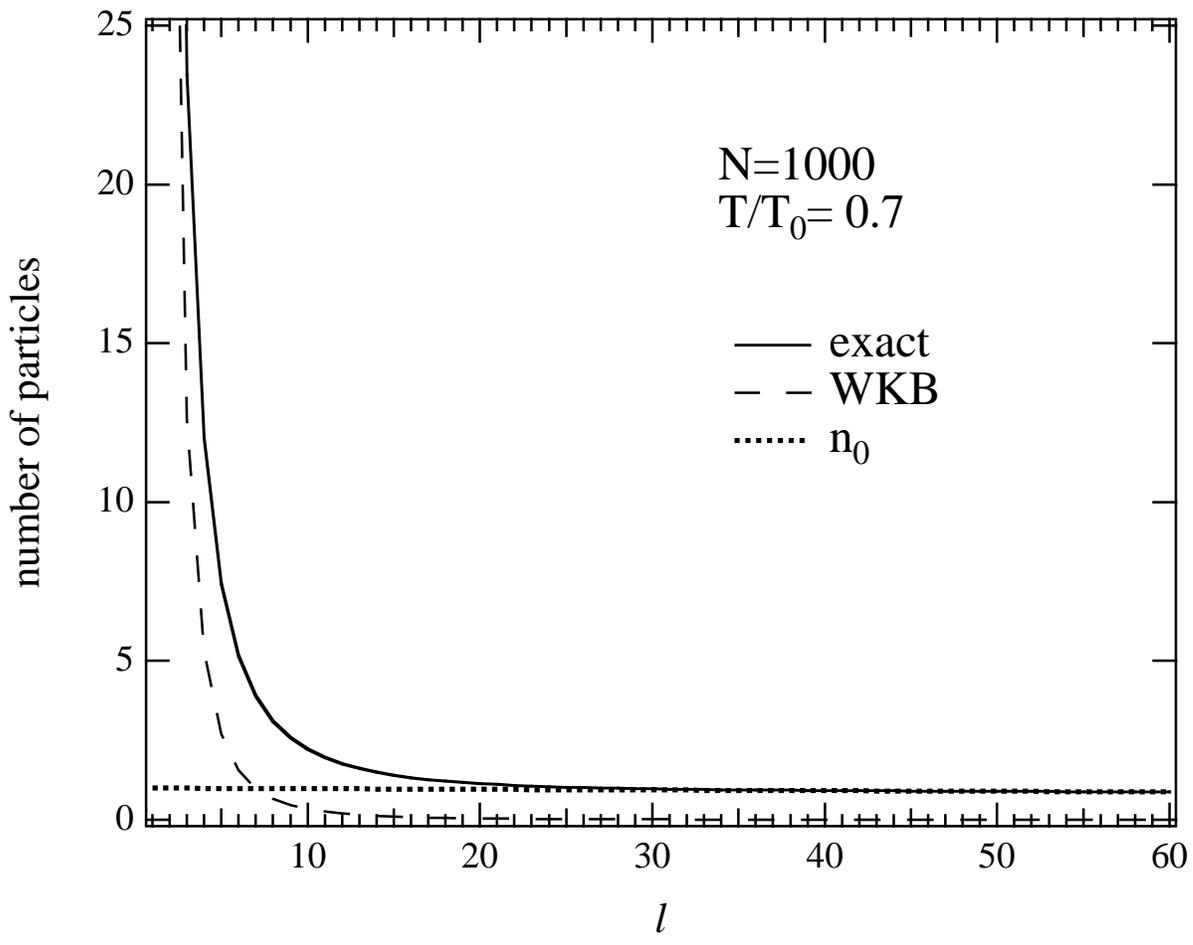
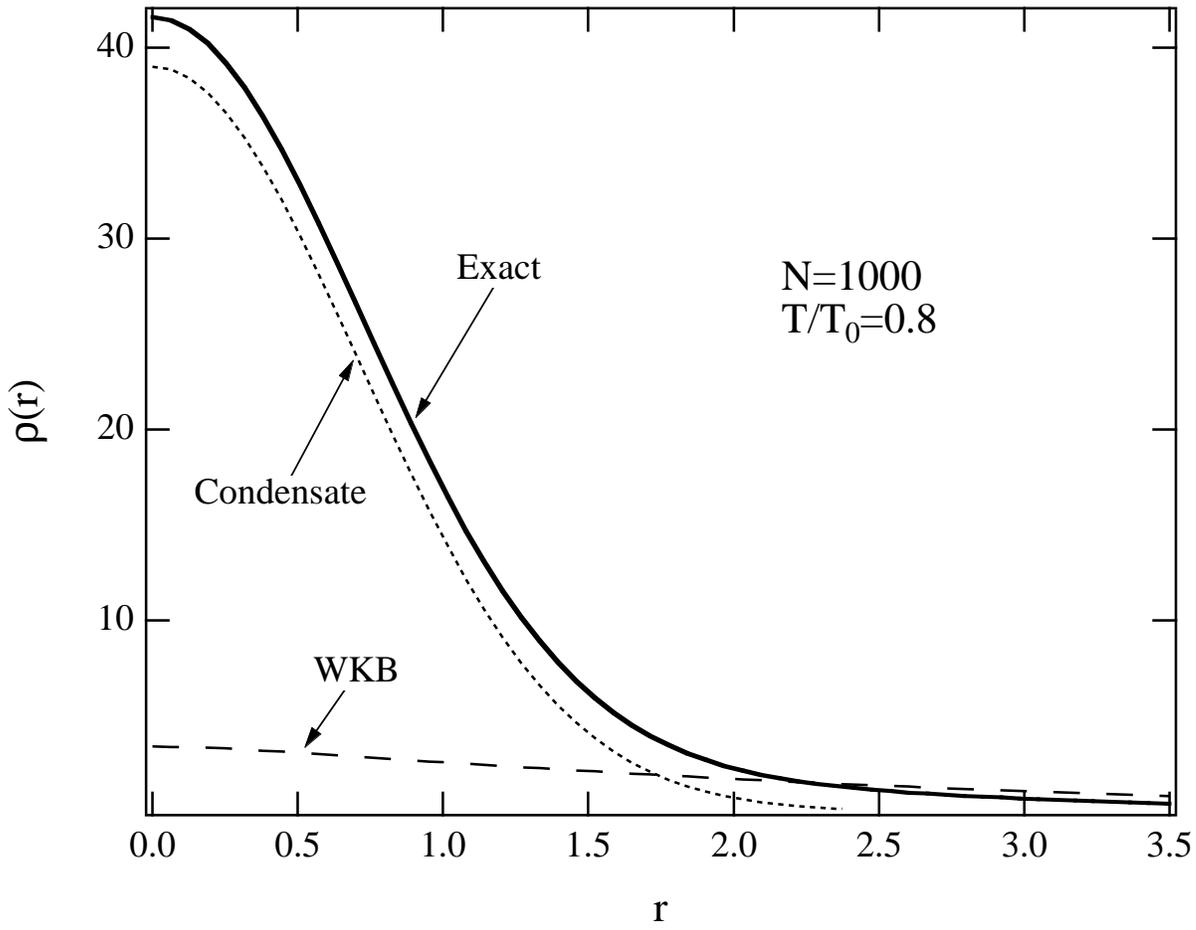
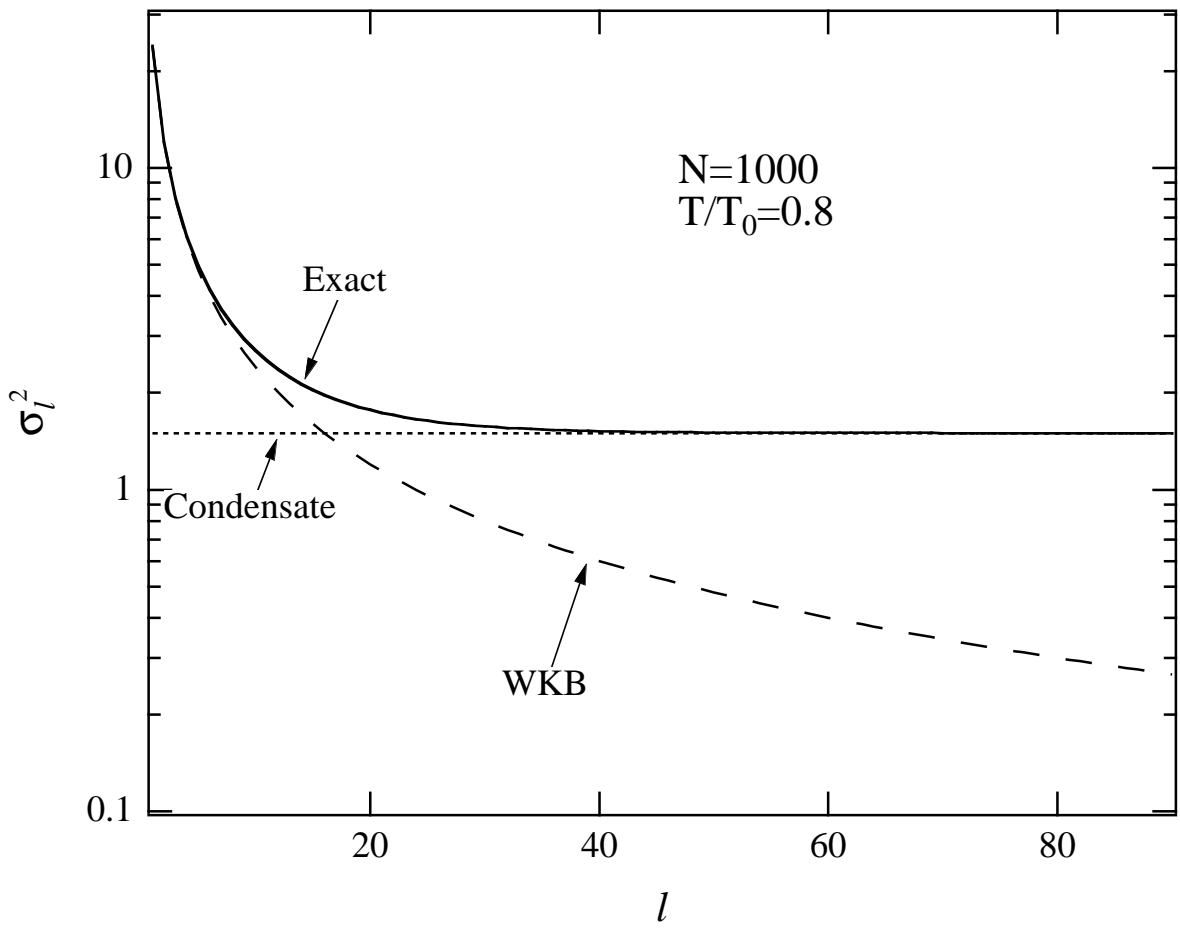
\end{document}